# A Review of Energy Efficient Dynamic Source Routing Protocol for Mobile Ad Hoc Networks

Baisakh
Department of Computer Science and Engineering
Sambalpur University Institute of Information Technology

## ABSTRACT
Dynamic Source Routing protocol (DSR) has been accepted itself as one of the distinguished and dominant routing protocols for Mobile Ad Hoc Networks (MANETs). From various performance analysis and results, it is shown that DSR has been an outstanding routing protocol that outperforms consistently than any other routing protocols. But it could not pervade the same place when the performance was considered in term of energy consumption at each node, energy consumption of the networks, energy consumption per successful packet transmission, and energy consumption of node due to different overhead. Because, DSR protocol does not take energy as a parameter into account at all. And as MANET is highly sensible towards the power related issues and energy consumption as it is operated by the battery with the limited sources, needed to be used efficiently, so that the lie time o the network can be prolonged and performance can be enhanced. This paper presents a comprehensive summery of different energy efficient protocols that are based on the basic Mechanism of DSR and enlightens the effort and commitment that has been made since last 10 year to turn the traditional DSR as energy efficient routing protocol.

## 1. INTRODUCTION
Ad Hoc Network is a multi-hop wireless networks which is consist of autonomous mobile nodes interconnected by means of wireless medium without having any fixed infrastructure. It's quick and easy deployment in a situation where its highly impossible to set up any fixed infrastructure networks, has increased the potential used in different applications in different critical scenarios. Such as battle fields, emergency disaster relief, conference and etc. A mobile ad hoc network [MANET] [1, 2, 3, 4, 7] can be characterized by the mobile nodes which have freedom to move at any direction and has the ability of self-configuring, self-maintaining and self-organizing themselves within the network by means of radio links without any fixed infrastructure like based station, fixed link, routers, and centralized servers. As in the network there is no base station or central coordinator exists, so the individual node plays the responsibility as a router during the communication has to be played by each and every node, participating in the network communication. Hence all the nodes are incorporated with a routing mechanism in order to transmit a data packet from source to destination. Nodes are operated by battery which is having limited capacity and they all suffer from severe battery consumption, especially when they participate for data communication for various sources and destinations. An uninterrupted data transmission from a particular source to destination requires a continual updating of path. If any moment path is not fond from source to destination, then Route Discovery Process has to be called. And multiple times route Discovery Process may introduce heavy power consumption. A number of routing approaches have been proposed to reduce various types of power consumption caused by various reasons in the wireless ad hoc network, which in result not only prolongs the life span of individual nodes but also reduces the network partition and enhances the performance of the network.

In fixed infrastructure wireless network [5] is a static network where its different components have to be set up permanently prior to the establishment of the communication. It takes not only huge time but also involves a huge cost for establishing the network.

The best example of a fixed infrastructure based network is Global system for mobile Communication (GSM) known as Second generation Mobile cellular System which is also a wireless network. In GSM, network architecture comprises several base transceiver stations (BTS) which are clustered and connected to a base station controller (BSC).Several BSC are connected to an Mobile Switching Center (MSC). The MSC has access to several data base, including Visiting Location Register (VLR), Home Location register (HLR).It is also responsible for establishing, managing and clearing connection as well as routing calls to proper radio cells. Here even if the nodes are mobile but they are limited with a fixed number of hops while communicating with other nodes.

But in case of MANET, it is completely different. The network is considered as a temporary network as it is meant for a specific purpose and for a certain period of time. And it is based on multi-hop technology where the data can be transmitted through number of intermediate nodes from source to destination.

With the rapid demands of MANET in the recent years, certainly have challenged the researchers to take up some of the crucial issues like bandwidth utilization, limited wireless transmission range, hidden terminal and exposed terminal problem, packet loss due to transmission error, mobility, stimulated change of route, security problem and battery constraint.

One of the important challenges of MANET is power constraint. The mobile ad hoc networks are operated on battery power. And the power usually gets consumed in mainly two ways. First one is due to transmitting of data to a desired recipient. Secondly, mobile node might offer itself as an intermediate forwarding node in the networks. The power level of the node is also getting affected while any route is established between two end points. The tradeoff between frequency of route update dissemination and battery power utilization is one of the major design issues of ad hoc network protocols. Because high power consumption will increase the battery depletion rate which in turn reduces the node's lie time, network lie time and causes network partition. Due to high network partition performance et a affected due to increase in number of retransmission, packet loss, higher end to end delay and many more problems.

Therefore, various energy efficient routing protocols have been proposed to increase the lifetime of the nodes as well as lifetime of the networks, so that communication can be carried out without any interruption. This article provides as well as analyzes different energy efficient routing protocols designed for ad hoc wireless networks which are only based on the mechanism of traditional DSR routing protocol.

The remaining of the session is organized as follow. The next section-2 presents two subdivision of Ad Hoc routing Protocols and their basic routing mechanism. We have basically emphasized the basic working principle o DSR routing protocol as it we have explained all energy efficient routing protocol which is based on DSR only. In section-3 we have shade some lights on the requirement of energy aware routing protocol for MANET and its different approaches to achieve that goal. The next section-4 highlights all related work that has been done to make DSR as an efficient energy aware routing Protocol. And finally the last section-5 concludes the article.

## 2. ROUTING PROCESS IN AD HOC NETWORKS

In MANET [1, 2, 3, 4, 6,7], routing is a process of establishing a route and then forwarding packets from source to destination through some inter mediate nodes if the destination node is not directly within the range of sender node. The route establishment itself is a two steps process. First one is the Route Discovery where it finds the different routes from same source to destination. Second, the Route Selection, where it selects a particular route among all routes found for the same source to destination. Traditional protocols and data structure are available to maintain the routes and to execute it by selecting the path that is having minimum distance from source to destination where the minimum distance is in term of minimum hop count.

### 2.1 Existing ad hoc routing protocols

Ever Since Defense Advance research Project (DARPA) sponsored packet radio networks in the early 1970, multifarious protocols have been developed for mobile ad hoc networks. And every protocol is having its own significance and meant for handling some of the issues like high data error rate, bandwidth utilization, security and high power consumption. As many routing algorithm were designed for dealing with these issues mentioned above, were expected to have certain characteristics fulfilling the minimum criteria that is needed for any mobile ad hoc networks. These characteristics are given below.

- It must be fully distribute, as centralized routing is less fault tolerant than the distributed routing which involves less risk single point of failure.
- It must be adopted to frequent topology change caused by the mobility of the node.
- Route maintenance must involve a minimum number of nodes.
- It must be loop free and stale route.
- It must converge the optimal route once the network topology becomes stable.
- Number of packet collision must be kept to a minimum by limiting the number of broadcast made by each node.
- It must use the scare resource such as bandwidth, computing power as demanded by the applications.

Considering all the above characteristics, the protocols have been divided into different classes where one protocol may fall under more than one classes. Depending upon the mechanisms, they have been classified. And so based on the routing information update mechanism the protocols are divided into two types.

- Table driven or Proactive Protocols
- On Demand or Reactive Protocols

### 2.1.1 Table driven or proactive protocols
Proactive Protocol as the name signifies, each node keeps all routing information to every other nodes in the network by maintaining one or more routing tables. These routing formations get updated periodically in the table to maintain the latest view of the networks. It comes in use when a node requires a path to a destination. Some of the existing table driven protocols are DSDV, DBF, GSR, WRP, ZRP and many more. This article does not cover all these table driven protocols as it is focused on DSR and different modification made on DSR protocols.

### 2.1.2 Reactive or on demand routing protocols
The protocols which fall under this category is completely different from the previous one. Here the protocols are On – Demand routing protocols [8, 12] do not update the routing information periodically as there is no routing table present for keeping routing information. Each node has route cache rather than routing table where it keeps all latest paths from source to destination. Rather a path is obtained when it is to establish a communication path between a source to a destination. Some of the example of on demand routing protocols are DSR, AODV, TORA, ABR etc.

The emphasis in this research paper is concentrated on survey of different energy efficient routing protocols based on the basic Mechanism of DSR only. The next subsection describes the basic features of DSR protocols.

## 2.2 Dynamic source routing protocol

Dynamic Source Routing (DSR) [9, 10] is a simple and efficient routing protocol designed specification for use in multi-hop wireless ad hoc mobile networks. DSR is one of the important routing protocols that are used for mobile ad hoc networks as much energy efficient routing protocols are designed based on its mechanism. It finds the route from source to destination only when the source initiates route discovery process. All aspects of protocol operate entirely on demand. This protocol also makes the network self-organizing and self configuring. Basically the protocol is composed of two mechanisms, Rote Discover and Route Maintenance and these two mechanisms work together to allow nodes to discover and maintain the source route to any destination node in the a hoc networks.

- Route Discovery
- Route Maintenance

### 2.2.1 Route discovery
Route discovery is done with two sub steps that is Route request and Route Reply.

#### 2.2.1.1 Route request
The route discovery comes in play when a mobile node has some data/packet to send to any destination and it does not have any route to the destination in its route cache. Then it initiates route discovery by broadcasting a **route request (RREQ) packet.** This route request contains address of the destination, address of the source and a unique identification number that is generated by the source node only. Each node receives the packet and checks whether the packet is meant for it or not. If it is not the destination node then it simply forwards the packet to the outgoing links adding its own address in the packet. To avoid duplicate route request which is generated from the same source, a node only forwards the route request that has not yet been seen appear in the route request with the same identification number.

#### 2.2.1.2 Route reply
As soon as the packet arrives at the destination node or arrives at a node that contains in its route cache an unexpired route to the destination, then a route reply is generated. Not only the packet contains all the address of the intermediate node it has come across but the sequences of hops are also stored in it. The Route reply is generated by the destination placing the route record contained in the route request into route reply.

During the route reply if the destination node has the route to the initiator in its route cache, It may use that route for route reply. Otherwise destination node may reverse the route in the route record if the link is symmetric. If the symmetric links are not supported then the node may initiate its own route discovery piggybacking the route reply on the new route request. When any intermediate node receives any route reply from destination node or any other node then they append their route record and forwards it to its neighbor nodes.

### 2.2.2 Route maintenance
Route maintenance is a process of identifying link whether it is reliable and capable of carrying packet on it or not. This process is executed by the use of route error packets and acknowledgements. When the data link layer encounters a **fatal transmission problem** then a route error message is generated. Suppose a packet is retransmitted (up to a maximum number of attempts) by some hop the maximum number of times and number of receipt conformation in received, then this node returns a packet error message to the original sender of the packet, identifying the link over which the packet could not be forwarded.

## 2.3 Benefits and Limitation
As the entire route is contained in the packet header, there is no need of having routing table to keep route for a given packets. The caching of any initiated or overheard routing data can significantly reduced the number of control message being sent, reducing overhead.

But DSR is not scalable to large networks. The internet draft acknowledges that the protocol assumes that the diameter of the network is not greater than10 hops. Additionally DSR requires significantly more processing resources than most of other protocols. The other drawback of DSR is selecting the path for routing on the basis of minimum hop counts from the source to the destination. As it selects the path of having minimum hops count, lesser will be the number of intermediate nodes, more will be the distance between each pair of nodes. As the distance is more we need to have more transmission power to communicate between any pair of nodes and hence it consumes more battery power as it is one of the limited resources.

## 3. ENERGY EFFICIENT ROUTING PROTOCOLS
The energy efficient routing protocols [6, 11] play a significant role in mobile ad hoc networks as the nodes are dynamic in nature and each node can participate in routing the data packets. In such scenario, efficient routing protocols are needed for Ad Hoc networks, especially when there are no routers, no base stations and no fixed infrastructure. So establishing the correct and efficient routes between the a source and destination is not the ultimate aim of any routing protocols, rather to keep the networks functioning as much as possible with less battery consumption at each node, should also be the objective of any routing protocols.

These goals can be accomplished by minimizing mobile node's energy during both the active as well as inactive communications. Active communication is when all the nodes of the route are participating in receiving and forwarding of data. Minimizing the energy during active communication is possible through two different approaches:

- Transmission power Control
- Load distribution

In an inactive communication the nodes are idle i.e. neither forwarding any data packets nor receiving any data packets. In such situation, to minimize the energy consumption Sleep/Power-down approach is used. We will not discuss about the power consumption during inactive communication in the network. There are many energy matrices used for calculating the power consumption caused by different reasons. The energy few energy related metrics are used. These metrics are helpful while determining energy efficient routing path instead of considering shortest path like in the traditional DSR protocol use. These metrics are:

- Energy consumed per packet
- Time to network partition
- Variation in node power level
- Cost per packet
- Maximum node cost

By using these metrics we can determine the overall energy consumption for delivering a packet, which is known as Link cost. In other word, link cost is the transmission energy over the link. Basically the efficient energy protocol selects the minimal power path depends which minimizes the sum of the link cost along the path.

## 3.1 Transmission Power Control Approach
We assume that a node's radio transmission power [13, 14] is controllable, if its direct communication range as well as the number of its intermediate neighbors is also adjustable. As the transmission power increases, the transmission range also increases and it reduces the number of hop count to the destination. Weaker transmission makes topology sparse and it may result more network partition and high end to end delay.

So it is desirable to have perfect transmission range between any pairs of nodes, so that less power consumption will occur. And it is possible when the transmission power can be adjustable according to the requirement of the receiver. So, instead of having high or low transmission power between the pair of nodes let the transmission power be set in such a way that any pair of nodes just reachable to each other. It will not only save the energy of battery but also reduces the interference and congestion in the networks.

## 3.2 Load Distribution Approach
The main objective of load distribution approach [16] is to select a route in such a way that the underutilized nodes will come in play rather than the shortest route. Due to the proper load distribution among the node, there is high balance in energy usage of all nodes. This approach certainly do not provide lowest energy route but surely prevent certain nodes from being overloaded and contributes towards longer network life time of the node.

## 3.3 Sleep/Power-down Approach
This approach is used during inactive communication. When any node is not receiving or transmitting any packets to other node, then it is desirable to put the subsystem/hardware into the sleep state or simply turn it off to save energy.

## 4. RELATED WORK
Since last 10 years many energy efficient routing protocols have been proposed and wondering the best solution out of all. As it is very difficult to restrict technologies and research digging for optimal solution, many noticeable enhancement and modifications have been done to convert DSR as an energy efficient routing protocol and serve it as efficient routing protocols like other protocols. So in the next session here are few important routing protocols which are made after doing some modification in traditional DSR protocol.

## 4.1 Global Energy Aware Routing (GEAR) Protocol
We have discussed that the Route Request is propagated towards the destination via multiple intermediate nodes. In Global Energy Aware (GEAR), along with the route request it piggybacks the remaining battery power as well as its identity and broadcast it to its neighbor nodes. When the destination nodes receive these different route request (RREQ) from the same source, then it selects the best route on the basis of high remaining/residual battery power out of the all received RREQ. But it does not guarantee the selected path is the best path always. Because it may happen that few routing paths with a better metric may not be considered, if it is arrived some time later than the specified time duration.

Apart from the above problem GEAR is associated with two major problems, one is incapable o utilizing the route cache and the other one is blocking property. Because the individual node is not having any power related information in its route cache which induces traffic surge due to the flood of the RREQ. Whereas the other problem is to manipulate the waiting time of the various RREQ from the same source node in order to select the best route by the destination node. Because prior to select the best route it waits until it receives all RREQ messages along all possible routing paths. So while specifying the time duration, it should be taken into consideration that too short time may not select the best path always. On the other hand too long time may affect the average response time. However local energy routing protocol comes as the remedy for the problems discussed above.

## 4.2 Local Energy Aware Routing (LEAR) Protocol

Localized Power Aware Routing (LEAR) Protocol [14] is based on DSR routing mechanism. The basic idea of LEAR is to consider only those nodes for the communication which are willing to participate in the routing path. This "Willingness" is the special type of parameter used in the modified DSR to find the route from source to destination. The new parameter can be determined by the Remaining Battery Power ($E_r$). If it is higher than a "threshold Value ($th_r$)", then the node will be considered for the route path and 'Route Request' is forwarded, otherwise the packet is dropped. It means only when the intermediate nodes will have good battery levels then only the destination will receive route request message. So the first message that arrives at the destination will be considered to follow an energy efficient as well as reasonably shortest path.

An interesting situation arises when a single intermediate node of a total route has lower battery power level than its threshold value(Er<Thr), a route request is simply dropped. If it is occurred for every possible path then the source will never receive a single route reply message even there exist a path between a source to destination. To avoid this situation, the source will resend the same route with an increased sequence number. When any intermediate node receives this same route request message, again with larger sequence number it adjusts (lower) threshold value to continue forwarding.

As the LEAR is mean to estimate the energy consumption and the balance across all mobile nodes, the result from the simulation shows that it has achieved the balanced energy consumption across all nodes successfully which is 35% more than that of DSR. Where as in the traditional DSR, energy consumption at different layer are uninformed, some nodes consume less energy and some nodes consume more energy. And hence the LEAR provides longer transmission time compared to DSR.

LEAR routing protocol not only achieves balanced energy consumption based only on local information but also removes the blocking property of GAER. Other than that it has also an advantage of being its simplicity characteristic and being integrated easily into existing ad hoc routing algorithm without affecting the other layers of protocol stack. This was the first work to explore the balanced energy consumption in the realistic environment taking DSR as its base protocol where mobility, radio propagation, routing algorithm are concerned. From the simulation it was shown that in LEAR there was better distribution of energy then the traditional DSR routing protocol.

## 4.3 Energy Saving Dynamic Source Routing (ESDSR) Protocol

Energy Saving Dynamic Source Routing (ESDSR) protocol [15] is another modified DSR protocol which is aimed to prolong the network life time by using basic two approaches of power consumption, one is transmission power control approach and the second one is load balancing approach. In the first phase it decides the route based on the load balancing approach and in the second phase it dynamically adjusts the transmitting power at every node before it transmits packet.

The idea came from the traditional routing mechanism which is basically based on minimum hop count. Instead o having minimum hop count approach while selecting the path and having fixed transmitting power, it introduces two new parameters. One is the current energy level and the other one is the current transmitting power level o individual node. Because it assumes that the ration of the current power level and the current transmitting power is nothing but the depletion rate o the battery. When the following cost is maximize then only a source node finds the route R(t):

$C(R,t) = \max ( R_j (t) )$ …….. (1)

$R_j (t) = \min ( E_i / P_{ti} )$ ………(2)

Where $R_j (t)$ is called the minimum expected lift time at time "t" or the path j. So while selecting the path it selects the path which is having maximum of minimum expected life time among different possible path. Then each node calculates the minimum transmitting power in order to send the packet to it's next neighbor node. This minimum transmitting power is calculated in the following way:

$P_{min} = P_{tx} + P_{recv} + P_{threshold}$ …(3)

Where $P_{tx}$ is the transmitting power to send the packet, $P_{recv}$ is the receiving power of the node at which it receives the packet, required threshold power of the receiving node for successful reception of data. Each node maintains a power table where the required transmitting power of that node and it transmits the packet at that power.

ESDSR is implemented by considering various parameters like total numbers of data packets reached at the destination, energy consumption per packet, number of dead nodes and outperforms better than DSR routing protocol irrespective these different parameters.

## 4.4 Energy Dependent DSR Routing (EEDSR) Protocol

Energy Dependent DSR (EEDSR) [16] is also an energy efficient routing protocol which is based on traditional DSR mechanism. It is almost similar to the LEAR approach but the only difference is that the willingness factor depends upon some other parameters. These parameters decide whether a node should participate in forwarding the packets or not which in turn it prevents nodes from a sharp drop of battery power.

The concept behind this algorithm is to compute the residual battery power (RBPi) of each node (ni) periodically. If the node has enough residual battery then it can participate in the network activities behaving exactly as DSR nodes. But when it's residual power becomes less than the specific threshold, the node delays broadcasting of a RREQ. As the delays in the node increases the **predicted lifetime** decreases.

### *4.4.1 Predicted lifetime*

The predicted lifetime simply denotes when the remaining battery of the node (ni) is exhausted. Predicted life time of node (ni) can be calculated by taking the ratio of Residual battery Power of the node and drain rate. Drain rate examines how much energy is consumed per second.

So with these above metrics, EEDSR attempts to discourage node which are having low residual power and high drain rate. So the node with small predicted life time will be rejected and cannot participate in forwarding the data. When energy of a node along an active route falls below a critical threshold, then it will immediately inform the source sending RREP packet. The source will try to find another route to the same destination by using Route Discovery Process.

In LEAR as the mechanism is based on residual battery power, much traffic load are injected through node if the node had high value of remaining power. And so drain rate becomes high in such nodes. but in EEDSR, the as route is obtained considering both the factors, the remaining battery power as well as the drain rate of the nodes, overcomes the problem of high traffic on the channel.

From the simulation it is found that EEDSR performs better than DSR as well as LEAR. In a dense networks scenario, EEDSR obtained a high number of survived nodes and improves Drain Rate in term of average node life time. Whereas the LEAR suffered from flooding problem caused by Drop_Route_Request packet sent in broadcast manner.

## **4.5 Energy Efficient DSR Protocol (E2DSR)**

The Energy efficient DSR (E2DSR) [17] is one of the splendid efforts made so far in order to make DSR as an efficient routing protocol. Because It has introduced many significant parameters as performance matrices which helps in calculating energy consumption in MANET. Even though it has the same objective like other protocols, but it has left a broad scope or the research activities. It is one of the latest energy aware routing protocols designed to reduce power consumption in battery effectively by doing some modifications in DSR. E2DSR has proposed some new structure for the control packets to change the behavior of the nodes implements a new Energy table and creates a new algorithm for route cache and route selection.

### *4.5.1 Structure of the Control Packets in E2DSR*

A new field is inserted in the RREQ message in the called Energy Field by using the data structure as an array. The energy field contains the remaining energy power of each node that is forwarded with the RREQ. This energy represented by some bits. If the number of bits are 4 then there can have 0 to 15 different energy level of the battery. The level 0 indicates the battery is empty and level 15 indicates the battery is full.

The structure of the RREP has also modified like the RREQ message. It appends the array of the energy level with the RREQ message which has to be received by the destination.

### *4.5.2 Routing Behavior of Nodes in E2DSR:*
### *4.5.2.1 Intermediate nodes*

When first RREQ arrives from source to any intermediate node, it stores the RREQ in the request table. Then extract the energy array of the every RREQ and calculates the energy parameters of the path. If the energy parameter of the first RREQ is E1 then it is added with the threshold value (E).

If the second RREQ, the value of E2 is less than e1, then that RREQ will be dropped. Note that, every intermediate node forwards and replies to a maximum of K RREQ from source to destination. The intermediate node will not forward request that have arrived after the T inter_wait Time expires.

### *4.5.2.2 Destination Node:*

The destination node has to play very simple role by replying the first received RREQ and also to the subsequent K-1 received RREQ which have the highest energy.

### *4.5.2.3 Source Node:*

In E2DSR, the source will select the best route for a specific destination by using a new function called Route priority Function. This route priority function will select the route with higher energy level. The process of selecting the route using Route priority Function is a background process which calculates the best route when any new RREQ is arrived at destination. The best route is chosen depending upon some new parameters like delay, jitter, and packet delivery ratio.

The route Priority function is basically depends upon three input parameters.

- Length
- Freshness
- Energy of path

**Length Parameter**: It is considered that the more will be the length of the link the higher probability of link breakage. So it is one of the important metric while evaluating energy saving in the networks. It can be calculated by the following method.

L(i)=Length of Route(i)/ Max_length of the route

**Freshness Parameter:** This parameter is somehow difficult to calculate due to dynamic nature of the ad Hoc Networks. Because sometimes the node's movement may change the validity of a route. So it must be required to know that which node is the fresh/latest and it is possible by keeping the information in the routing cache. The freshness of route (i) can be calculated as $F(i)=(n-i+1)/n$ for n number of routes. The freshest route has the freshness value 1.

**Energy Parameter**: Since there is high probability of network failure with the low battery nodes, such routes are undesirable. It can be evaluate by,

$E(i)=[(RE(i)-MRE(i)).MRE(i)]/M(i).Initial Energy^2$

Where $M(i)$= Number of nodes in route (i)

Initial energy = the maximum energy,

$RE(i)$=Total remaining energy in route( i),

$MRE(i)$=Minimum of the remaining energy between all nodes for route( i)

### 4.5.3 Performance metrics
As in E2DSR we consider different performance metrics to determine its performance with respect to the existing energy aware routing protocols which are based on DSR.

#### 4.5.3.1 Average End to End Delay:
It is used to determine the number of successfully received the 'n' packets sent by the source to the destination.

Average end to End Delay=$\sum_{j}^{n}(End\_time_j-Start\_time_j)/n$

#### 4.5.3.2 Inter arrival Jitter:
It determines how stable the routes in the Energy Efficient DSR are. It can be defined as the mean deviation of difference (D) in the packet spacing at receiver compared to the sender for a pair of packets.

It can be calculated as $D(I,j)= mod |(R_j-S_j) –(R_i-S_i)|$

Where $S_i$=Time Stamp from packet i, and $R_i$ = time of arrival for packet i.

#### 4.5.3.3 Normalized Routing Load (NRL):
It helps to determine the overhead incurs in E2DSR.NRL is the number of routing packets transmitted per data packet successfully delivered at the destination.

NRL = (Number of E2DSR packet sent during T)/n

Where "n" is the stream of n packets successfully delivered.

#### 4.5.3.4 Balancing of battery Power consumption:
It measures the effectiveness of the Energy balancing E2DSR algorithm. The smaller the deviation the more effective is energy balancing. It can be determined by using the following formula:

$EL(i)$ = [Consumed energy]/ [Total consumed energy]

#### 4.5.3.5 Nodes failure degree:
This parameter determines the capability of keeping nodes alive for more time.

Node Failure Degree = ((Number of failed nodes in T) / n Where n=Number of nodes in topology and T is the time window.

The simulation result shows that E2DSR has better performance than DSR. And it has left many doors opened for future studies and research works because all these metrics described above, have not been implemented yet for a large scenario. So we may get many interesting results while carry out any experiment with a larger scenario with all these new performance matrices.

## 4.6 Topology Control Based Power-Aware and Battery Life-aware DSR (TPBDSR) Protocol
It is realized later that the topology control has serious effects on the system performance in various ways. It can affect the traffic carrying capacity as well as can have the contention for the medium. Topology Control Based Power-aware Battery Life-aware DSR (TPBDSR) [18] uses simple pure distributed control where each node adjust its transmitting power through certain range of neighbor that are given with some number. If the node find it's neighbor within or beyond certain range then the transmitting power is getting adjusted. In other word, we can say that the transmitting power gets adjusted according to their neighbor node's position in the network topology which may change dynamically. This strategy also limit the power adjustment period which is denoted by h second, where the value of h may vary with mobility character of the networks.

### 4.6.1 The Operation in Intermediate nodes:
As TPBDDSR has to adjust the transmitting power, it requires Transmitting Power Value Field (TTP) and Least Battery Value field (LBP) which have to be attached with the Route request Packet. Every time the node k receives the route request packet (RREQ) ,then the new value of TTP has to be updated.

So, $TTP_{New} = TTP_{old} + TTP_k$ and if $B_k <$ LBPold then LBPnew = $B_k$

Where $B_k = F_k/ R_k(t)$,

$F_k$ = Full battery Capacity of the node, $R_k$ = remaining Battery capacity of the node k at time t.

### 4.6.2 The Operation in the Destination Node:
DSR basically accepts the first packet that has come through the shortest path and hence it discards other subsequent request packets. But TPBDSR accepts multiple route requests. It does that with some additional rules. The moment first route request arrives at the destination; it sets a timer and wait for a more route request packets containing other path's information. And selects the best path among all the paths it has in list.

Simulation result shows that the TPBDDSR have much longer life time of nodes than the traditional DSR and EEDSR. As the scenarios of the experiment changes the result may vary, hence it's still difficult to say which one is best between TPBDDSR and EEDSR. But certainly TPBDSR gives better results than DSR.

## 4.7 Minimum Energy Dynamic Source Routing Protocol (MEDSR)

Minimum Energy Dynamic Source Routing Protocol (MEDSR) [19] has done one of the finest attempts to make DSR more as an energy aware routing protocol. The whole MEDSR approach is based on two mechanisms:

- Route Discovery
- Link Power Adjustment

The route Discovery process itself is classified into two sub processes.

- Route Discovery mechanism using low power level
- Route discovery mechanism using high power level

### *4.7.1 Route Discovery mechanism using low power level:*
In this process of route discovery when source node S has some packets to send, then it sets a minimum level transmitting power for all the nodes. So the route packet will be broadcasted to only re within the range of the minimum level of transmitting power. Once the route request arrives at the destination, the destination node copies the power level information from the route request packet into the route reply. The route reply is sent back to the nodes that are within the small range of transmitting power level from the destination node. The moment, the intermediate node will receive the route reply, it will calculate the minimum power for itself. The minimum transmitting power level for any node can be calculated as

$P_{min} = P_{tx} - P_{rec} + P_{th}$

Where $P_{tx}$=Transmitting Power of Destination

$P_{rec}$=Receiving Power of the node that has received th route reply

$P_{th}$= Threshold receive power for successful reception of the packet. And it will keep continuing at each node until the route reply is received by the source. Once the route reply reaches at the source, the source sets the transmitting power. And start sending with that transmitting power on the route that is been selected for data transmission.

### *4.7.2 Route Discovery mechanism using High Power Level:*
High power route discovery is just same as the low power route discovery. The only difference is that instead of setting up the low transmitting power, it sets high transmitting power while sending route request. This process is highly needed for route discovery, especially when no path is found due to unreachability by setting the transmitting power low. So to overcome this problem high power routing is also mandatory.

MEDSR uses two levels of powers; the network connectivity is highly maintained and results less network partition. The result also depicts that when the network size is small the energy saving per data is maximum in MEDSR as compared to DSR, almost 55% high which indeed turning out to be an efficient routing protocol.

## 4.8 Modified DSR Routing (MDSR) Protocol

As the name suggests Modified DSR (MDSR) [20] has been designed after making some modification in DSR routing protocol. Because, due to source routing nature of DSR, the overhead increases when the network size gets increased. Apart from that, the energy consumption of nodes also increases as the nodes act as intermediate nodes for multiple sources destination pairs. So MDSR protocol has aimed to reduce overhead by reducing the number of routing reply packets and a fixed header size for the data packets and acts itself as one of the energy efficient routing protocol.

The main drawback of DSR protocol is the generation of redundant route replies. These RREP travels through the path in which RREQ's bearing same Ids were received. This causes not only congestion but also wastage of battery power.

So it's whole operation consists of two basic mechanisms:

- Overhead reduction mechanism
- Efficient energy management mechanism

### *4.8.1 Overhead Reduction Mechanism:*
#### *4.8.1.1 Control Packet Reduction:*
In MDSR, the destination alone initiates a RREP only for the first received RREQ rather than for every RREQ bearing the same id which is reached via different routes.

#### *4.8.1.2 Fixed header size in MDSR*
Unlike DSR, MDSR makes the address part of the packet header to carry only the source and destination address. It offers to traverse through a correct path from source to destination even if the intermediate nodes are not added with the packet. It does it by creating a **routing table** which stores source and destination address instead of having **route cache.**

#### *4.8.1.3 Efficient Energy management mechanism:*
In the traditional DSR, the transmitting power is set same for all the nodes which are participating in packet forwarding process. But in MDSR, the transmission power varies between a pair of nodes depending upon the distance between them.

To calculate the Minimum required transmission power for each node, the two Ray Model is used.

$P_t(min) = [(P_r(th) \times d^4)/(h_t^2 \cdot h_r^2 \cdot G_t \cdot G_r)] + C$

13

The RREP packets help the intermediate nodes to calculate the distance between itself and neighboring node; so that the required transmit power can be determined. And this required transmitting power is stored in the routing table along with the source address, destination address and precursor node address before the RREP is broadcasted. Upon receiving the route reply the source starts transmitting the packet with the required transmitting power.

The simulation results that MDSR has generated less number of control packets than the existing DSR. Here control packets are sum of all the RREQs, RREPs and Route Error Packet. MDSR has also high packet delivery ratio as compared to the existing DSR. Due to the reduction of number packets and dynamic variation of transmit power, provides high saving of energy than the traditional DSR.

## 4.9 Multi-path Energy Aware DSR Routing (MEADSR) protocol

By the time, energy efficient protocols were becoming the interest of all researchers in Mobile Ad Hoc Network, at the same time some other issues were started evolving as major constraints of MANET. Due to self organizing and self configuring nature of MANET, it generates lot of control packets and because of wireless in nature, these all were the major cause for high traffic overhead and high bandwidth consumption. So multi path routing was introduced as one of the remedy of these problems. As multipath routing [22] reduce the number of route discovery and reduce the end to end delay time, bandwidth utilization became fair. But the multi path routing was highly involved in generating control messages. To make it more efficient one, both the multipath and energy aware techniques are integrated called as Multi–path Energy Aware DSR (MEADSR) [21] routing protocol. So the basic objective of this protocol is to have the best path for routing by computing multipath node disjoint where the best path is the high energy efficient. MEADSR performs its operation by two major steps:

- Node disjoint path discovery
- Update mechanism

### 4.9.1.1 Node disjoint path discovery:
Traditional DSR stores all paths that the RREP have travelled through. But in MEADSR, those paths are stored which are having node-disjoint routes. Because path-disjoint-ness inside multi-path DSR can increase data throughput by handling the greater resilience against the node mobility.

$$P1 - (s,d) \cap P2 - (s,d) = \emptyset$$

Where P1 and P2 are two different paths are considered disjoints whose intersection is empty. These paths are ordered not just by the path length but by an energetic metric. This energetic metric is calculated while the route reply comes across the intermediate nodes from destination to the source.

### 4.9.1.2 Update Mechanism:
Since the active node which are participating in the communication, continuously receive and transmit the packet, they consume their energy. This energy is calculated while RREP is travelled through these intermediate nodes. So the change in energy should be reflected at the source route cache which contains the information about the path length as well as the cost unction o the entire path <Path, Ci>. MEADR uses a special packet called RREQ-PROBE which is a uncast probe packet sent by all nodes. This packet is a kind of RREQ packet and the moment destination node will find the probe packet it will send the RREP-PROBE packet and the updated Ci values are stored here within the packet. The moment source will receive the RREP-Probe packet it just updates its route cache.

MEADSR results prove that there is evenly distribution of energy consumption among nodes by using their residual battery capacity. And it is considered as one of the best energy efficient routing protocol which is made by modifying the traditional DSR routing protocol.

## 5. Conclusion

In this article we have discussed, one of the important issue that is energy consumption problem in MANET. As the traditional routing mechanisms like minimum hop count produces not only overheads in the networks but also consume more power in the networks during the communication..And hence it is required to have some any energy efficient routing protocols to be designed in order to overcome this problem. As there are many energy efficient routing protocols exist, it is very difficult to compare them directly since each method has different assumptions and has different means to achieve the goals. For example to reduce the power consumption we can go for dynamic adjustment of transmitting power at each node. But due to mobility it may suffer from network partition as any node moves away from another node. If we consider the traffic density or node density in the network which are also responsible of power consumption in a node, can be solved by using distributed load balancing approach in the networks. DSR routing protocol has provided the basic for any energy efficient routing protocols where by modifying its structure of control packet and considering some new energy matrices power consumption can be reduced.

This paper has also expounded few energy efficient routing protocols which are explicitly based on the DSR routing protocol. These protocols have proved the traditional DSR can also be acted as an energy efficient routing protocol. Because DSR is considered as one of the unconventional routing protocol which does not concerned about energy consumption at all. This paper has also revealed that a single routing protocol cannot stand strongly against the major constraint of MANET that is power consumption until it is integrated with some other techniques like power consumption, load balancing, transmission control, multi path routing and many more. The combination of all these techniques can surely turning out be an efficient solution for energy constraint.

There can also be many variegated solutions and techniques proposed to deal with some other problems and constraints of MANET but power constraint cannot be ignored as long as MANETs are operated by the battery power.

It is very difficult to conclude which one of the protocol is the best among all energy efficient routing protocols, because all these protocols are based on different methodologies, performances matrices, different implementation environments and different techniques. But all these protocols have proved that they are better than the DSR routing protocol. Still many scopes are there in DSR to add on new functionally and to modify the basic mechanism of DSR as an Energy Efficient Routing protocols.